\documentclass[aps,pra,twocolumn,groupedaddress]{revtex4-1}

\usepackage{amsmath,amssymb,amsfonts}
\usepackage{graphicx}
\usepackage{siunitx}
\usepackage{natbib}
\usepackage{braket}
\usepackage{multirow}
\usepackage[normalem]{ulem}
\usepackage{verbatim}

\usepackage{url}
\usepackage[breaklinks]{hyperref}
\hypersetup{
    unicode=true,
    colorlinks=true,
    linkcolor=blue,
    citecolor=blue,
    urlcolor=blue
  }
	\usepackage{breakurl}
	\usepackage{xcolor}	

\def\txr{\textrm}
\def\mb{\mathbf}

\begin{document}

\title{Dynamics of a degenerate Cs-Yb mixture with attractive interspecies interactions}

\author{Kali E. Wilson$^1$}
\thanks{These two authors contributed equally.}
\author{Alexander Guttridge$^1$}
\thanks{These two authors contributed equally.}
\author{I-Kang Liu$^2$}
\author{Jack Segal$^1$}
\author{Thomas~P.~Billam$^2$}
\author{Nick G. Parker$^2$}
\author{N. P. Proukakis$^2$}
\author{Simon L. Cornish$^1$}
\email{s.l.cornish@durham.ac.uk}
\affiliation{$^1$Joint Quantum Centre Durham-Newcastle, Department of Physics, Durham University, South Road, Durham, DH1 3LE, United Kingdom.}
\affiliation{$^2$Joint Quantum Centre Durham–Newcastle, School of Mathematics, Statistics and Physics,
Newcastle University, Newcastle upon Tyne, NE1 7RU, United Kingdom}

\begin{abstract}
We probe the collective dynamics of a quantum degenerate Bose-Bose mixture of Cs and $^{174}$Yb with attractive interspecies interactions. Specifically, we excite vertical center of mass oscillations of the Cs condensate, and observe significant damping for the Cs dipole mode, due to the rapid transfer of energy to the larger Yb component, and the ensuing acoustic dissipation. Numerical simulations based on coupled Gross-Pitaevskii equations provide excellent agreement, and additionally reveal the possibility of late-time revivals (beating) which are found to be highly sensitive to the Cs and Yb atom number combinations. By further tuning the interaction strength of Cs using a broad Feshbach resonance, we explore the stability of the degenerate mixture, and observe collapse of the Cs condensate mediated by the attractive Cs-Yb interaction when $a_{\rm{Cs}}<50 \, a_0$, well above the single-species collapse threshold, in good agreement with simulations.
\end{abstract}

\date{\today}

\maketitle
Superfluid mixtures formed of ultracold atoms offer a versatile playground for studies of binary fluid dynamics in both Bose-Bose and Bose-Fermi mixtures. Interplay between superfluids can lead to counterflow instabilities~\cite{Takeuchi2010,Takeuchi2010b,Engels2011, Delehaye2015}, and quantum turbulence~\cite{Takeuchi2010,Norrie2006}. Previous work has explored collective excitations in the form of scissors mode oscillations in a repulsive Bose-Bose mixture~\cite{Modugno2002a}, and dipolar oscillations~\cite{Ferlaino2003}, breathing dynamics~\cite{Ferrier2014}, and the exchange of angular momentum~\cite{Roy2017} in Bose-Fermi mixtures.  Experimental control in ultracold atomic mixtures allows the balance of the inter- and intraspecies interactions to be tuned, enabling further manipulation of the coupled dynamics and stability of the system.  This can result in novel bound states such as bright-dark solitons~\cite{Tylutki2016}, and exotic vortex lattices~\cite{Yao2016, Mingarelli2019}, and enable miscibility studies \cite{Lee2016}. Prior research on dynamical instabilities associated with attractive interspecies interactions has focused on Bose-Fermi mixtures, where the bosonic species mediates collapse and subsequent density-dependent atom loss in the fermionic species~\cite{Ospelkaus2006b, Modugno2002, DeSalvo2019}. However, recent studies have used Bose-Bose mixtures to probe the region where the net mean-field interaction is close to zero, resulting in the observation of quantum droplets where quantum fluctuations provide a repulsive force that stabilizes against collapse~\cite{Cabrera2018, Semeghini2018, DErrico2019}.\\

In this Letter, we probe the collective dynamics and stability of a quantum degenerate Bose-Bose mixture of Cs and $^{174}$Yb with an attractive interspecies interaction characterised by the scattering length $a_\mathrm{CsYb} = -75 \, a_0$~\cite{Guttridge2018a}. The Cs intraspecies scattering length $a_\mathrm{Cs}$ is a tunable experimental parameter via a broad, low-field magnetic Feshbach resonance~\cite{Chin2004,Berninger2013}, whilst the $^{174}$Yb intraspecies scattering length is $a_\mathrm{Yb} = 105 \, a_0$.
We use vertical Cs center-of-mass (CoM) oscillations to probe the collective dynamics of the system, observing an upwards mean-field shift of the CoM dipole mode frequency. We also find that the attractive interspecies interaction results in complicated coupled dynamics, as Cs pulls along Yb in the region of Cs--Yb overlap. We probe the stability of the mixture by tuning the Cs interaction strength across the collapse threshold, demonstrating Yb-mediated collapse of the Cs Bose-Einstein condensate (BEC), well above the single-species collapse threshold.

\begin{figure}[t]
		\includegraphics[width=0.95\linewidth]{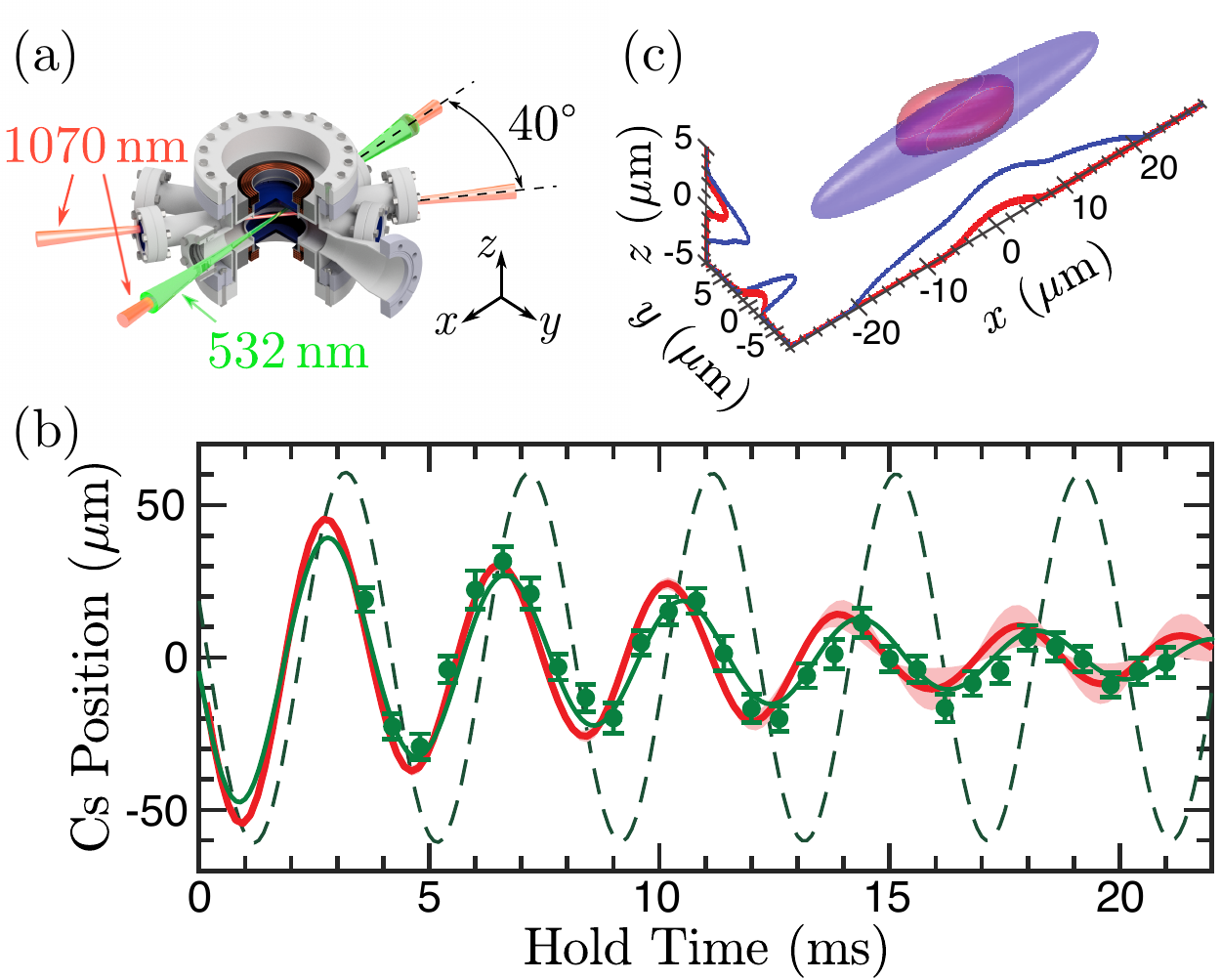}
\caption{Cs and Yb coupled dynamics. (a) Layout of the bichromatic optical dipole trap. (b) Vertical CoM position of the Cs BEC after 40\,ms ToF versus hold time in the trap measured experimentally (green points) and simulated numerically (red line).  The solid green line is a fit to the experiment: $A e^{-\beta t} \sin (\omega t + \phi_0)$  with $A = 52(8)\,\mu$m, $\omega = 2\pi \times 260(2)$\,Hz, and $\beta = 0.10(2)\,\mathrm{ms}^{-1}$. The dashed green line shows an equivalent fit to a measurement performed in the absence of Yb, yielding $\omega = 2\pi \times 251(1)$\,Hz~\cite{supplement}. The numerical plot is based on an average over 206 particle number sets, with the red band indicating the standard deviation. Atom numbers are randomly selected such that their mean is $(N_\textrm{Yb},N_\textrm{Cs})=(40,6.5)\times10^3$, with a 20\% standard deviation. 
(c) Numerically-generated coupled 3D density profiles for Cs (red) and Yb (blue) with $(N_\textrm{Yb},N_\textrm{Cs})=(40,6.5)\times10^3$. Solid lines are sliced density profiles along the $x$, $y$ and $z$ axes.
	\label{fig:TrapFreq}}
\end{figure}

We prepare dual-degenerate mixtures of $^{133}$Cs in the $\ket{F=3, m_{F}=+3}$ hyperfine state and $^{174}$Yb using the apparatus and procedures described in Refs.~\cite{Hopkins2016,Kemp2016,Guttridge2016,Guttridge2017,Wilson2020}. Our dual-species evaporation relies on sympathetic cooling of Cs with Yb, resulting in an imbalanced mixture with typical condensate atom numbers of $(N_\mathrm{Yb}, N_\mathrm{Cs}) = (50,5) \times 10^3$ with up to $20\%$ shot-to-shot variation~\cite{Wilson2020}. The condensates are confined in a bichromatic optical dipole trap (BODT), using laser beams with wavelengths of $532 \, \mathrm{nm}$ and $1070 \, \mathrm{nm}$ aligned as shown in Fig.~\ref{fig:TrapFreq}(a). We define a coordinate system $(x,y,z)$ with $x$ along the copropagating BODT beams and $z$ the vertical direction. The axes $\Tilde{{x}}$ and $\Tilde{{y}}$ for the Cs trap are rotated 20$^\circ$ from the Yb trap axes.  The final trap frequencies are $(\omega_\mathrm{x},\omega_\mathrm{y},\omega_\mathrm{z})_{\mathrm{Yb}} = 2\pi \times (10,120,80) \, \mathrm{Hz}$ for Yb and $(\omega_{\Tilde{\mathrm{x}}},\omega_{\Tilde{\mathrm{y}}},\omega_\mathrm{z})_{\mathrm{Cs}} = 2\pi \times (40,70,260) \, \mathrm{Hz}$ for Cs. The vertical trap frequencies are very sensitive to the overlap of the BODT beams and can vary by up to 10\% with day-to-day realignment of the trap~\cite{Wilson2020}. 

We first explore the coupled dynamics of the mixture by exciting Cs CoM oscillations in the vertical direction.  
At the end of the dual-species evaporation sequence we adiabatically ramp on a magnetic field gradient of 9.0\,G/cm in the $z$-direction.
This additional magnetic potential causes a $0.96(7) \, \mu$m vertical shift of the Cs trap minimum, without directly acting on Yb which has no magnetic moment. 
The gradient is rapidly switched off, exciting the vertical CoM dipole mode. We wait a variable hold time before turning off the trapping potential, then perform absorption imaging of the Cs condensate after 40\,ms of levitated time of flight (ToF).  We hold $a_\mathrm{Cs}$ constant at $275 \, a_{0}$ for both in trap oscillations and the subsequent ToF. To account for the high sensitivity of the vertical trap frequencies to alignment, we reference the coupled Cs--Yb dynamics to the Cs CoM oscillations in the absence of Yb by taking both measurements in quick succession. We obtain the Cs-only reference by selectively removing Yb with a 5\,ms pulse of light resonant with the $^{1}S_{0}\rightarrow{^{1}P}_{1}$ transition at $399 \, \mathrm{nm}$ prior to exciting the Cs CoM oscillations. 

The measured vertical CoM oscillations of the Cs BEC in the presence of the Yb BEC are shown by the green circles in Fig.~\ref{fig:TrapFreq}(b). These oscillations are strikingly different from the corresponding Cs-only oscillations (dashed green line), which occur at the Cs natural trap frequency $\omega_\mathrm{Cs,z} = 2\pi \times 251(1)$\,Hz.  In the presence of Yb, the Cs CoM oscillations experience a frequency shift upwards of $4(1)\%$, with significant damping. This shift is in good agreement with a simple mean-field prediction that assumes a Thomas-Fermi profile for the Yb BEC and neglects the back-action of the Cs BEC on the Yb atoms due to the large number imbalance~\cite{supplement}. Under such assumptions, the interspecies interaction results in an additional attractive quadratic potential for Cs that leads to a $5\%$ upwards frequency shift. 

We perform numerical simulations based on coupled 3D Gross-Pitaevskii equations (GPE)~\cite{supplement} to gain further insight into the rich oscillatory dynamics.  A numerical visualization of the initial geometry with $(N_\mathrm{Yb},\,N_\mathrm{Cs})=(40,\,6.5)\times10^3$ atoms, $\omega_\mathrm{Cs,z} = 2\pi \times 250 \, \mathrm{Hz}$,  and an imposed $+0.95 \, \mu\mathrm{m}$ vertical shift of the Cs trap position is shown in Fig.~\ref{fig:TrapFreq}(c). The numerical simulations reveal that both components start the coupled dynamics from non-equilibrium positions with respect to their trap centers.  The attractive Cs--Yb interaction induces a density `bulge' in the Yb BEC in the region of Cs--Yb overlap, visible in the $z$-axis  density profile shown in Fig.~\ref{fig:TrapFreq}(c). The off-center bulge leads to a vertical Yb CoM shift upwards of $\Delta z_\mathrm{Yb,CoM} = 0.1 \, \mu\mathrm{m}$. Similarly, the Cs CoM shift reduces to $\Delta z_\mathrm{Cs,CoM} = 0.88 \, \mu\mathrm{m}$.  

The simulated vertical CoM oscillations of the Cs BEC are shown by the red line in Fig.~\ref{fig:TrapFreq}(b). The simulations indicate that the CoM shifts and ensuing coupled dynamics are highly sensitive to the relative atom numbers~\cite{supplement}. Therefore, to account for shot-to-shot variations in both Cs and Yb atom numbers, the simulation result is based on the average of 206 atom number combinations, randomly selected such that their mean and standard deviations are $(N_\mathrm{Yb},\,N_\mathrm{Cs})=(40,\,6.5)\times10^3$ and 20\%, respectively, in broad agreement with the values measured experimentally. As shown in Fig.~\ref{fig:TrapFreq}(b), this leads to excellent agreement with the experimental observations. A video of the coupled dynamics is available in the Supplemental Materials (SM) \cite{supplement}.
Closer analysis of the simulations reveals the dynamics are complicated by the different vertical trap frequencies experienced by the two components $(\omega_\mathrm{Cs,z} \approx 3 \, \omega_\mathrm{Yb,z})$, the large atom number imbalance 
$(N_\mathrm{Yb}/N_\mathrm{Cs} \approx 6)$, and the broader spatial extent of the Yb BEC in the weakly-confined direction.  Accordingly, in what follows we focus our analysis on the central bulge in the Yb BEC that directly interacts with the Cs BEC. 

\begin{figure}[t]
    \centering
    \includegraphics[width=0.95\linewidth]{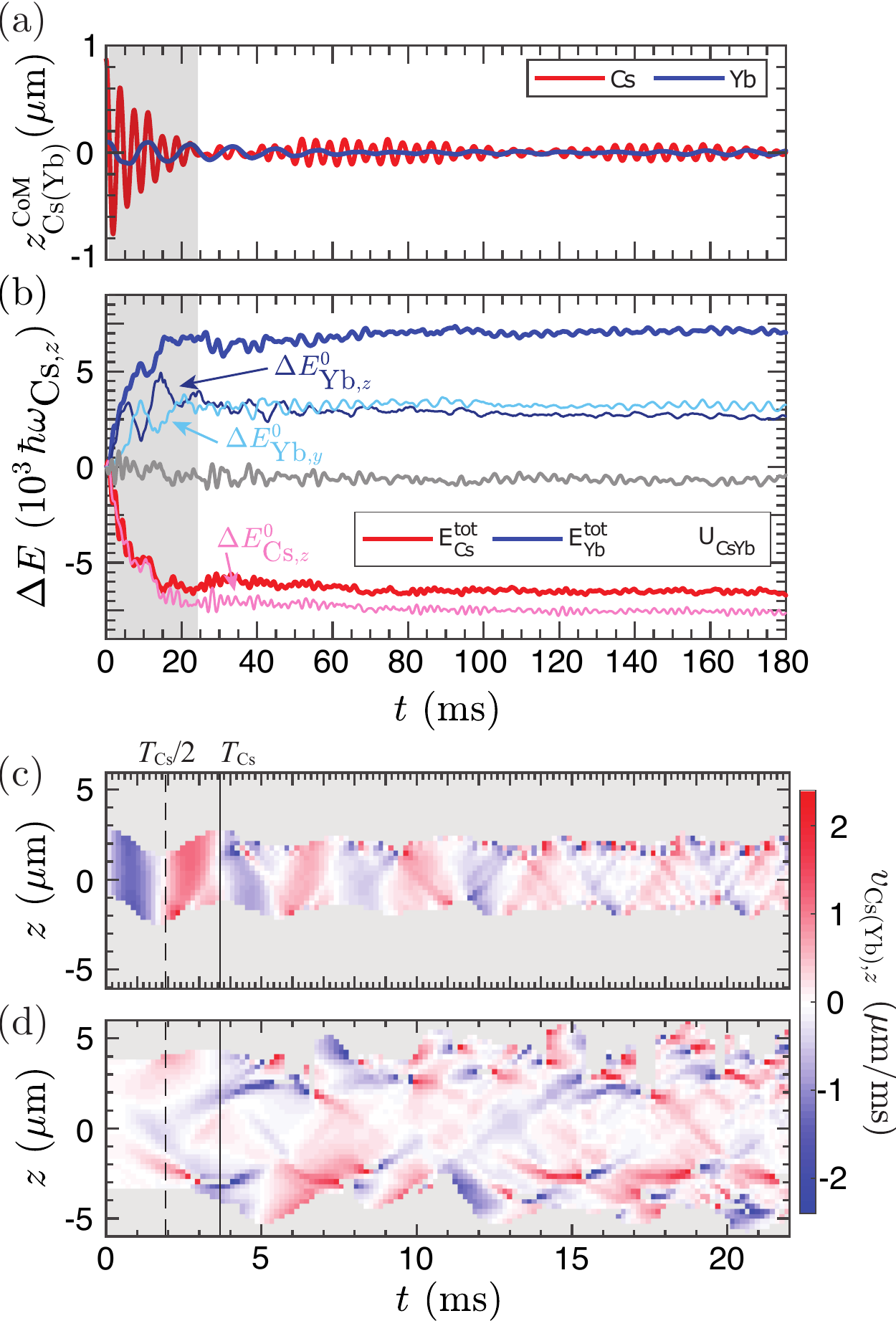}
    \caption{Cs and Yb excitation and energy exchange calculated numerically in a coupled-GPE treatment with $(N_\textrm{Yb},N_\textrm{Cs})=(40,6.5)\times10^3$. 
    (a) Extended numerical evolution of CoM motions. (b) Total energies of Cs and Yb ($E_\textrm{Cs/Yb}^{\textrm{tot}}$), the interspecies interactions energy ($U_{\textrm{CsYb}}$), and the main corresponding single-particle energy changes along the $z$ and (for Yb only) the $y$ axis. Note the more gradual initial increase along $y$, consistent with a secondary excitation mechanism, and the initial out-of-phase oscillations between $z$ and $y$ directions for Yb. (c) Time evolution of the Cs velocity profile $v_\mathrm{Cs}=\hbar\nabla\arg[\psi_\mathrm{Cs}(0,0,z)]/m_\mathrm{Cs}$. (d) Corresponding Yb velocity profile. Velocity plots include a grey mask to filter out regions corresponding to atomic densities lower than $2\,\mu\mathrm{m}^{-3}$.
    \label{fig:GPE}}
\end{figure}

A striking feature of Fig.~\ref{fig:TrapFreq}(b) is the fast damping of the Cs CoM oscillations in the presence the Yb BEC, with the amplitude falling to $1/e$ its initial value in 10(2)\,ms. We explore the origin of this damping in Fig.~\ref{fig:GPE}. Figure~\ref{fig:GPE}(a) shows the numerical CoM motions out to 180\,ms, whilst Fig.~\ref{fig:GPE}(b) shows the time evolution of the relevant energies. Looking first at the energies in Fig.~\ref{fig:GPE}(b), we see that there is a net loss of $\sim 85 \%$ of the initial excess Cs potential energy over the first $\sim 20$\,ms. This energy is transferred to the Yb component through a coupling of the CoM motions to low-lying excitations. Although the initial excitations are produced along the $z$ axis of relative motion, an additional secondary mode coupling between the $z$ and $y$ directions gradually develops owing to the comparable Yb trap frequencies in these directions. This is evident in Fig.~\ref{fig:GPE}(b) as an out-of-phase oscillation in $E_\mathrm{Yb,y}$ and $E_\mathrm{Yb,z}$, correlated with oscillations of the condensate widths. The end result is that after about $20$\,ms,
the initial excess Cs single-particle energy concentrated along $z$ has been converted to Yb single-particle energy across both the $z$ and $y$ directions in comparable fractions. Smaller, secondary couplings along the $x$ direction (the weak axis of the trap) occur at longer timescales~\cite{supplement}.

Earlier related studies of repulsively-coupled nonlinear superfluids~\cite{Abad2015,Ferrier2014,Castin2015,Laurent2019} 
have identified two distinct excitation mechanisms. Firstly, in a number-imbalanced system like ours, the motion of the smaller superfluid through the larger component can lead to single-component excitations in the larger superfluid.  For the attractive interaction in our system, this is analogous to stirring a single-component superfluid with a focused red-detuned laser beam~\cite{Weimer2015,Singh2016}. Secondly, hydrodynamic considerations in the homogeneous limit reveal the existence of a dynamical instability  whenever the relative motion of two superfluids exceeds some characteristic speed. In the limit of small interspecies interactions, this speed approaches the sum of the sound velocities of the two (uncoupled) superfluids. This counterflow instability leads to the simultaneous generation of pairs of excitations in both superfluids~\cite{Abad2015}. We believe both these mechanisms are at play in our experiment~\cite{supplement}.

The temporal evolution of the local velocity fields shown in Fig.~\ref{fig:GPE}(c) for Cs and Fig.~\ref{fig:GPE}(d) for Yb is useful for visualizing the two excitation regimes.  The velocity fields along the $z$ axis for each component $\alpha$ = \{Cs, Yb\} are obtained from $v_{\alpha,\mathrm{z}}=\hbar\nabla \arg[\psi_{\alpha}(0,0,z)]/m_\alpha$, where $m$ is the mass and $\psi$ is the condensate wave function. 
During the initial half oscillation ($0 < t < T_\mathrm{Cs}/2$), Cs exhibits coherent harmonic motion, maintaining a constant phase profile across its vertical width. In contrast, during this same initial period, incoherent features gradually emerge in the Yb component in the form of counter-propagating density waves as the relative velocity between the two components increases, consistent with the single-excitation picture. Here the Cs BEC is both smaller than Yb, and executes faster oscillations. Therefore, it acts as a perturbing attractive barrier, which travels through the Yb BEC, and drags along the central Yb density bulge. As shown in Fig.~\ref{fig:GPE}(d), the bulge begins to track the downward Cs motion at $t \sim 1$\,ms ($T_\mathrm{Cs}/4$), which corresponds to the point of maximum downward Cs velocity. This in turn causes Cs to slow down as it continuously transfers energy to Yb.

Counterflow dynamics occur when Cs reverses its motion at $t\sim2$\,ms ($T_\mathrm{Cs}/2$) whilst the central Yb bulge continues its downward motion, thus creating a direct counterflow between the two components, whose magnitude increases over the next quarter-cycle of the Cs oscillation (until $t = 3T_\mathrm{Cs}/4$).  
This leads to the onset of counterflow instabilities, which are particularly pronounced at the edges of the mixture's central overlap region ($|z| \gtrsim 2 \, \mu$m), where the relative speed exceeds the sum of the local critical velocities of the two components.
Moreover, the moving Cs component induces a secondary, at times opposing in direction, flow within the central Yb bulge, thus giving rise to the co-existence of flows in the same and opposing directions (the extension of co-moving and counter-moving modes for two superfluids which are not in the same trap \cite{Lee2016}). The ensuing Yb motion, which extends beyond the region of CsYb overlap, feeds back into Cs by exciting phonons at the edges of the Cs component where its density is lowest~\cite{supplement}, leading to a complicated energy transfer and phonon excitation mechanism in both components.
We have also verified that smaller initial amplitude oscillations give rise to reduced damping and fewer excitations. 

Although the dominant dynamics occur during the experimentally probed timescale of $22 \, \mathrm{ms}$, closer inspection of the long-time coupled evolution for a fixed total atom number reveals smaller scale energy transfers between the Cs and Yb components. This results in partial revivals in the Cs CoM oscillations, as shown in Fig.~\ref{fig:GPE}(a).
Such beating arises from the existence of two closely-located dominant frequencies for Cs (both within a few $\%$ of the natural Cs frequency), the origin of which can be attributed to the co-existing secondary dynamical micromotions (combinations of appropriate co-moving and counter-moving modes) in the central $z$ region. The combination of rapid damping and frequency sensitivity to particle number combinations make it hard to characterize such secondary effects in our current experiments. More details of these complicated dynamics are given in the SM~\cite{supplement}.

\begin{figure}
		\includegraphics[width=0.95\linewidth]{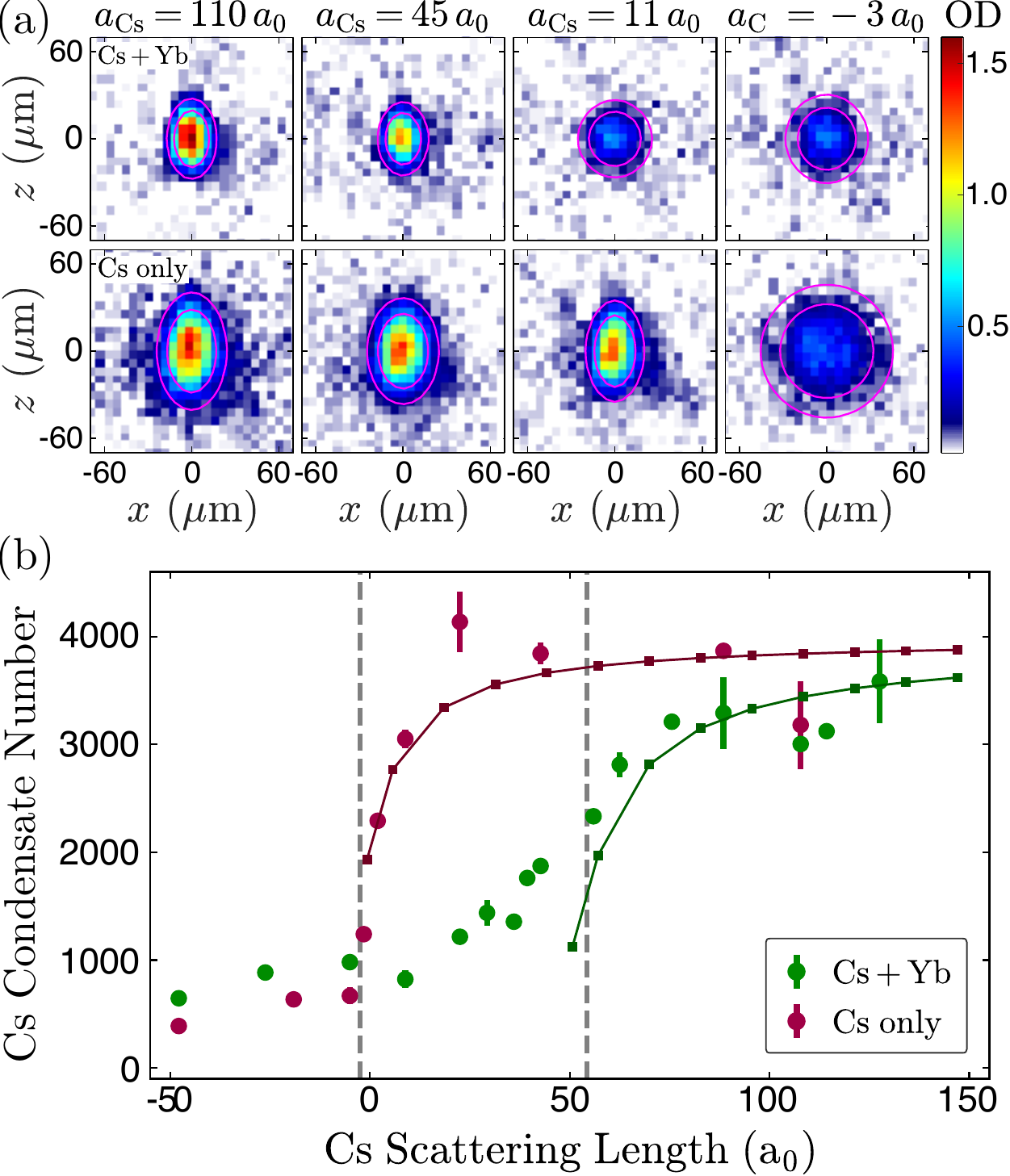}
\caption{Dual-species collapse. (a) Cs collapse mediated by $^{174}$Yb (top row) and Cs single-species collapse (bottom row). Images show the Cs BEC optical depth (OD) profile for varying $a_\mathrm{Cs}$. Pink contours show $1/e$ and $1/e^2$ of peak OD.  (b) Cs condensate number as a function of Cs scattering length for Cs collapse mediated by $^{174}$Yb (green circles) and Cs single-species collapse (red circles).  Dashed line at $a_\mathrm{Cs} = -2.5 \, a_0$ denotes the theoretically predicted single-species collapse threshold for $N_\mathrm{Cs} = 4 \times 10^3$. Dashed line at $a_\mathrm{Cs} = 54.3 \, a_0$ corresponds to the cancellation of mean-field interactions $\delta g = 0$. The results of GPE simulations with three-body loss are shown with green (Cs+Yb) and red squares (Cs only); lines are guides to the eye.
	\label{fig:Collapse}}
\end{figure}

In a separate study, we explore the dynamical instabilities of the mixture further by tuning the relative mean-field contributions through changes to the Cs scattering length, $a_{\mathrm{Cs}}$. The attractive interspecies interaction between Cs and $^{174}$Yb gives rise to mean-field collapse of the Cs BEC mediated by the presence of the Yb BEC, as shown in Fig.~\ref{fig:Collapse}. In both single-component and dual-species BECs, the onset of collapse is related to a mechanical instability arising from the population of imaginary Bogoliubov modes~\cite{Pethick2002}.  In a single-component BEC, collapse~\cite{Sackett1999,Gerton2000,Roberts2001,Donley2001,Cornish2006}, along with its associated density-dependent atom loss, occurs when the intraspecies interaction becomes sufficiently attractive to overcome the zero-point kinetic energy associated with the harmonic trapping potential. For a given atom number $N$, the scattering length $a_\mathrm{crit}$ associated with the onset of collapse is given by $|a_\mathrm{crit}| = C\sqrt{\hbar/(m\bar{\omega})}/N$,
where $\bar{\omega}$ is the geometric mean of the trap frequencies, and $C$ is a numerical constant weakly dependent on the trap geometry~\cite{Roberts2001,Ruprecht1995,Gammal2001}. 
In contrast, the dual-species collapse threshold primarily depends on the balance of the inter- and intraspecies interactions, and the critical point for collapse is determined by the parameter $\delta g = g_{12} + \sqrt{g_{11} + g_{22}}$ which describes the balance of mean-field interactions in the system.  Here, $g_{ij} = {2\pi\hbar^2 a_{ij} (m_i + m_j)} / (m_i m_j)$ is the interaction coupling constant. 
When $\delta g < 0$, the modes of the density branch of the Bogoliubov excitation spectrum become imaginary, leading to the onset of mechanical instabilities, and subsequent collapse~\cite{Ospelkaus2006b, Modugno2002, DeSalvo2019}.  

In our experiment, we approach the collapse instability by varying $a_\mathrm{Cs}$ via a magnetic Feshbach resonance. We begin with a dual-species BEC formed with $a_\mathrm{Cs} = 147 \, a_0$, with atom numbers $(N_\mathrm{Yb}, N_\mathrm{Cs}) = (50, 4) \times 10^3$.  We ramp $a_\mathrm{Cs}$ to the desired value over 10\,ms, hold for 30\,ms, and then ramp back to $a_\mathrm{Cs} = 147 \, a_0$ over 10\,ms prior to ToF expansion and imaging. As shown in Fig.~\ref{fig:Collapse}, we observe the onset of collapse as a marked decrease in Cs atom number due to density dependent 3-body losses ($K_{3,\mathrm{Cs}} \sim \SI{1e-27}{cm^6 s^{-1}}$~\cite{Guttridge2017,Kraemer2006} for the range of magnetic fields explored).  Figure~\ref{fig:Collapse}(a) shows absorption images of the Cs BECs for varying Cs scattering lengths, both with (top row) and without (bottom row) Yb present.  The presence of Yb causes the collapse instability to occur at $a_\mathrm{Cs} > 0$, where we would expect a stable single-component Cs BEC. In addition to atom loss, we also observe a change in aspect ratio for the Cs cloud.  We quantify the collapse in Fig.~\ref{fig:Collapse}(b) by plotting Cs condensate number versus Cs scattering length.  Red circles denote Cs single-species collapse occurring when $a_\mathrm{Cs} \leq 0$.  The vertical line at $a_\mathrm{Cs} = -2.5 \, a_0$ marks the predicted collapse point $a_\mathrm{crit}$ for $N_\mathrm{Cs} = 4 \times 10^3$ atoms.  Green circles show Cs collapse mediated by the presence of Yb.  The vertical line at $a_\mathrm{Cs} = 54.3 \, a_0$ marks the point where $\delta g = 0$ for the CsYb mixture. 

We model the experimental protocol numerically using two-component coupled GPEs with a density dependent 3-body loss term for Cs, and $\omega_\mathrm{Cs,z} = 2\pi \times 260$\,Hz~\cite{supplement}. The simulation results are shown in Fig.~\ref{fig:Collapse}(b) for Cs+Yb (green squares) and Cs only (red squares). For Cs scattering lengths lower than the range shown, the simulations do not converge at the highest resolutions we could feasibly use, indicative of mean-field collapse~\cite{supplement}. We find good agreement with the experimental results using a thermal Cs three-body loss coefficient of $K_{3,\mathrm{Cs}} = \SI{1.8e-27}{cm^6 s^{-1}}$.  However, we note that the Yb-mediated collapse is broader than predicted by the GPE simulations. This may be indicative of beyond-mean-field effects, such as the formation of quantum droplets recently observed in both K spin mixtures~\cite{Cabrera2018,Semeghini2018} and Rb--K mixtures~\cite{DErrico2019}, and is a direction for future studies.

In conclusion, we have probed the collective dynamics of a quantum degenerate Bose-Bose mixture of Cs and $^{174}$Yb with attractive interspecies interactions. Observations of the CoM oscillations of Cs in the presence of Yb have revealed a measurable increase in the dipole mode frequency and significant damping as energy is transferred to Yb. We find good agreement with numerical simulations which highlight the complexity of the coupled dynamics.
We have also investigated the stability of the degenerate mixture by crossing the transition for dual-species collapse and find good agreement with coupled-GPE simulations that include a Cs three-body loss term. 
This opens up new prospects for the study of the BEC to quantum droplet phase transition.
In this context, a key advantage of Cs--Yb mixtures is that the Cs scattering length may be precisely tuned without affecting Yb.
Additionally, the very different atomic structure of Cs and Yb enables species-specific optical traps with low photon scattering rates.
These advantages may prove invaluable for precise studies of binary fluid dynamics
in both Bose-Bose and Bose-Fermi Cs--Yb mixtures.

\begin{acknowledgments}
We acknowledge support from the UK Engineering and Physical Sciences Research Council (grant numbers EP/P01058X/1, EP/T015241/1 and EP/T01573X/1). IKL and NPP acknowledge funding from the Quantera ERA-NET cofund project NAQUAS (EPSRC EP/R043434/1). The data presented in this paper are available from
\url{http://dx.doi.org/}.
\end{acknowledgments}

%

\newpage
\renewcommand\thefigure{SM.\arabic{figure}}
\setcounter{figure}{0} 
\begin{center}
{\bf Supplemental Materials}
\end{center}

In these Supplemental Materials we present details of our numerical modelling, offering further evidence to support the findings made in the main text.

We start by describing our theoretical model which is based on a three-dimensional mean-field description by the coupled Gross-Pitaevskii equations~(GPEs),
\begin{equation}\begin{array}{rl}
\displaystyle i\hbar\frac{\partial\psi_\textrm{Yb}}{\partial t}=&\displaystyle\Big[-\frac{\hbar^2\nabla^2}{2m_\textrm{Yb}}+V_\textrm{Yb}(\mathbf{r})+\frac{4\pi\hbar^2a_\textrm{Yb}}{m_\textrm{Yb}}|\psi_\textrm{Yb}|^2
\\\\
&\displaystyle\;+\frac{4\pi\hbar^2a_\textrm{CsYb}}{M_\mu}|\psi_\textrm{Cs}|^2-\mu_\textrm{Yb}\Big]\psi_\textrm{Yb},
\end{array}
\label{eq:GPE_Yb}
\end{equation}
and
\begin{equation}\begin{array}{rl}
\displaystyle i\hbar\frac{\partial\psi_\textrm{Cs}}{\partial t}=&\displaystyle\Big[-\frac{\hbar^2\nabla^2}{2m_\textrm{Cs}}+V_\textrm{Cs}(\mathbf{r})+\frac{4\pi\hbar^2a_\textrm{Cs}}{m_\textrm{Cs}}|\psi_\textrm{Cs}|^2
\\\\
&\displaystyle\;+\frac{4\pi\hbar^2a_\textrm{CsYb}}{M_\mu}|\psi_\textrm{Yb}|^2-\mu_\textrm{Cs}-\frac{i\hbar K_{3,\textrm{Cs}}^{(\txr{C})}}{2}\Big]\psi_\textrm{Cs},
\end{array}
\label{eq:GPE_Cs}
\end{equation}
where $\psi_\textrm{Yb(Cs)}$ is the mean-field wavefunction of Yb (Cs) trapped in the harmonic potential, $V_\textrm{Yb(Cs)}(\mathbf{r})$ and $M_\mu=2m_\textrm{Yb}m_\textrm{Cs}/(m_\textrm{Yb}+m_\textrm{Cs})$ is the reduced mass.
The Cs three-body loss coefficient for the condensate is $K_{3,\textrm{Cs}}^{(\txr{C})}$ while the corresponding thermal coefficient $K_{3,\textrm{Cs}}$ is a factor of $3!$ higher~\cite{Burt1997}.
The wavefunctions are normalized to the total atom numbers,
\begin{equation}
\displaystyle\int d\mathbf{r}|\psi_\textrm{Yb}|^2=N_\textrm{Yb} \, , \quad \mathrm{and} \quad
\displaystyle\int d\mathbf{r}|\psi_\textrm{Cs}|^2=N_\textrm{Cs}\,.
\label{eq:N_CsYb}
\end{equation}
The trapping potentials are considered as
\begin{equation}
V_\textrm{Yb}=\frac{m_\textrm{Yb}}{2}\left(\omega_{\textrm{Yb},x}^2\,x^2+\omega_{\textrm{Yb},y}^2\,y^2+\omega_{\textrm{Yb},z}^2\,z^2\right)
\label{eq:V_Yb}
\end{equation}
and
\begin{equation}
V_\textrm{Cs}=\frac{m_\textrm{Cs}}{2}\left[\omega_{\textrm{Cs},x}^2\,\tilde{x}^2+\omega_{\textrm{Cs},y}^2\,\tilde{y}^2+\omega_{\textrm{Cs},z}^2\,(z-\Delta z)^2\right]
\label{eq:V_Cs}
\end{equation}
where the `tilde' notation $\tilde{x}$ and $\tilde{y}$ axes in $V_\txr{Cs}$ indicates that the longitudinal ($x$) and one of the transverse ($y$) axes of Cs are rotated by $20$ degrees with respect to the $x$ and $y$ axes of Yb while $z$ axis remains coincided, in agreement with the experimental setup, and $\Delta z$ is the trap displacement in Cs.

In the following, we investigate the coupled oscillations and collapse via GPE modeling. The GPEs are simulated by the Fourier pseudo-spectral methods with the Runge-Kutta method. In the simulations, the trapping frequencies are as described in the experimental section and we take $a_\textrm{Yb}=105\,a_0$ and $a_\textrm{YbCs}=-75\,a_0$, where $a_0$ is the Bohr radius.
In the coupled oscillation section, we consider fixed $a_\textrm{Cs}=275\,a_0$ with time-dependent $\Delta z$ and ignore three-body loss, whereas in the collapse section, $a_\textrm{Cs}$ is considered as a time-dependent variable with $\Delta z=0$, and we include the Cs condensate three-body loss term.

\section{Coupled Oscillations}
The coupled oscillations of Cs and Yb BECs triggered by the sudden removal of the trap displacement in Cs, are modeled by the GPEs without the inclusion of the three-body loss term, based on an initial Cs trap displacement $\Delta z = 0.95\;\mu$m, and  $\omega_{\textrm{Cs},z}=2\pi\times250$~Hz.
The initial trap displacement in Cs is implemented in the stationary initial condition, generated by solving Eq.~(\ref{eq:GPE_Cs}) with the imaginary-time propagation method. Subsequently, $\Delta z$ is set to zero for all $t\geq0$ to mimic the experimental procedure.
The required set of particle numbers for the initial conditions is reached by a self-consistent choice of the chemical potentials $\mu_\textrm{Cs(Yb)}$~\cite{Modugno2003} during the imaginary-time evolution. 

\begin{figure}[t!]
\centering
\includegraphics[width=1\linewidth]{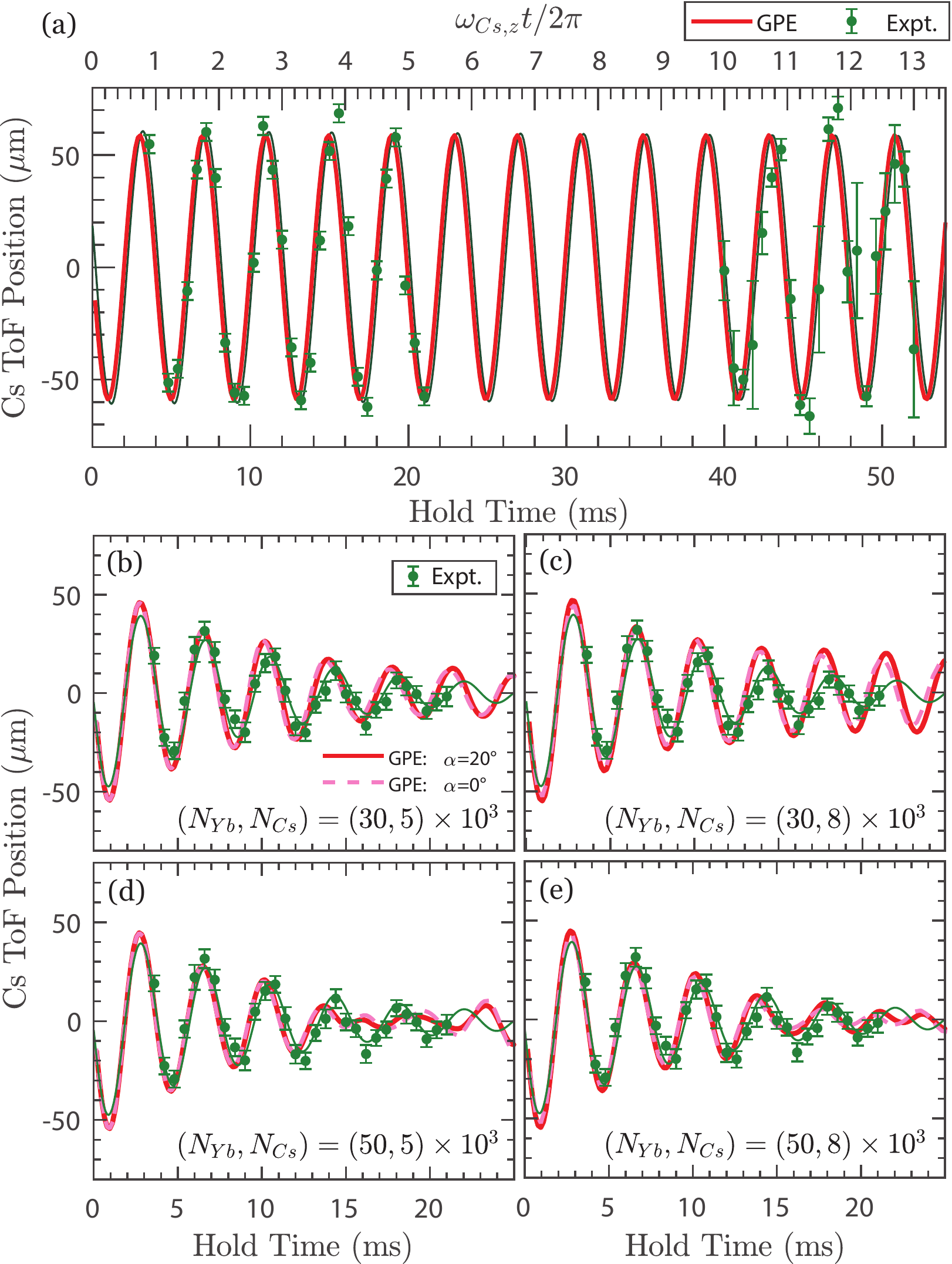}
\caption{ToF Cs CoM positions along the $z$ direction predicted numerically (red line) and measured experimentally (green points) (a) in the absence and (b)-(e) in the presence of the Yb component. Numerical data in (a) using the 'primary' Cs atom number $N_{\rm Cs}=6.5\times10^3$.
The experimental data set in the presence of Yb are compared to simulations corresponding to different atom numbers, consistent with atom number variations of up to 1.25$\sigma$ from the `primary' values $(N_\textrm{Yb},N_\textrm{Cs})=(40,6.5)\times10^3$.
In addition to numerical predictions in the actual experimental geometry with $\alpha=20^\circ$ (red solid lines), 
subplots (b)-(e) also show corresponding dynamics for $\alpha=0^\circ$
(dashed pink lines). The green line shown is the corresponding best fit to the experimental dataset.
}
\label{fig_expt_vs_GPE}
\end{figure}

\subsection{Time-of-Flight Analysis}

In the experiments, the center of mass (CoM) motion is measured by time-of-flight~(ToF) imaging.
Since a full ToF simulation is numerically forbidden due to the huge computational size constraints imposed by the $V_\textrm{Cs(Yb)}$ anisotropy, our numerical simulations focus on the characterization of the {\it{in situ}} CoMs, with an effective mapping to the ToF measurement along the kick axis.
The {\it{in situ}} CoM along the $z$ direction is defined by
\begin{equation}
z_{\textrm{Cs(Yb)}}^{\textrm{{CoM}}}=\frac{1}{N_\textrm{Cs(Yb)}}\int d\mathbf{r}\,z\,|\psi_\textrm{Cs(Yb)}(\mathbf{r})|^2\;.
\end{equation}

The presence of the attractive Cs--Yb interaction implies that the actual initial shift exhibited by the CoM of Cs is reduced, while simultaneously the Yb density is locally distorted by Cs, acquiring an initial CoM shift.
All the results presented here are based on a definite particle number combination.
For our `primary' atom number combination of $(N_\textrm{Yb},N_\textrm{Cs})=(40,6.5)\times10^3$ we find such initial shifts to be $z_\textrm{Cs}^{\textrm{CoM}}(t=0) \sim 0.88\,\mu$m, in comparison to the imposed trap shift ($\Delta z = 0.95\,\mu$m), and $z_\textrm{Yb}^{\textrm{CoM}}(t=0) \sim 0.10\,\mu$m. This `primary' atom number combination is in broad agreement with the values measured experimentally, and leads to excellent agreement with the experimental observations [see Fig.~1(b) of the main text].

Direct comparison to the experimental ToF data is facilitated by mapping the numerical {\it{in-situ}} results to the predicted ToF measurement by
\begin{equation}
\textrm{Cs ToF Position}\approx\tau_\textrm{ToF}\,\frac{dz_{\textrm{Cs}}^{\textrm{CoM}}(t)}{dt},
\label{eq:ToF}
\end{equation}
based on an expansion time $\tau_\textrm{ToF}=40$~ms, and on the assumption that the condensate carries its moving velocity during the ToF expansion.
The time derivative is computed by the central finite difference method, namely, $dz_\textrm{Cs}^{\textrm{CoM}}(t)/dt\approx[z_\textrm{Cs}^{\textrm{CoM}}(t+0.5\Delta t)-z_\textrm{Cs}^{\textrm{CoM}}(t-0.5\Delta t)]/\Delta t$, with $\Delta t=0.5/\omega_{{\rm Cs},z}$. This mapping provides an excellent agreement of numerical predictions with experimental observations as shown in Fig.~\ref{fig_expt_vs_GPE}.

Specifically, Fig.~\ref{fig_expt_vs_GPE}(a) shows the Cs CoM oscillation in the absence of the Yb component both based on experimental measurements (green points/line) and corresponding numerical predictions (red line), which yields (atom-number-independent) oscillations at the expected trap frequency $\omega_{{\rm Cs},z}$. 
The corresponding comparison for the Cs CoM oscillations in the presence of Yb is shown in  Fig.~\ref{fig_expt_vs_GPE}~(b)-(e). Specifically, the same set of experimental data (green points/line) are compared against numerical simulations (red lines) using 4 different atom number combinations, chosen here to reflect number variations of up to $\approx1.25\sigma$ in either Cs ($N_{\rm Cs} = (6.5 \pm 1.5) \times 10^3$) or Yb ($N_{\rm Yb} = (40 \pm 10) \times 10^3$). These reveal very good overall agreement over the range of available experimental data, but with a sensitivity of the late-time [$t>(15-20)$~ms] oscillations to the precise atom numbers (and in particular to the Yb atom number, which is the largest component).
[Recall that Fig.~1(b) of the main paper is based on weighted combinations of different atom numbers, statistically distributed around the `primary' mean value.]

The effect of the tilted axes in Cs ($\alpha = 20^{\circ}$) is also investigated (dashed lines in Fig.~\ref{fig_expt_vs_GPE}~(b)-(e)), and it is found to have a marginal influence on the CoM motion (at least over the probed timescales), clearly ruling out such axes mismatch as a primary reason for the observed dynamics. 

\subsection{Excitation Dynamics Analysis}

Here we analyze the dominant collective mode frequencies, provide information regarding the numerically-observed excitations and explain the physical origin leading to the beating at longer times.

\subsubsection{Excitation Profile \label{Sec_excitations} }

\begin{figure*}[t!]
\centering
\includegraphics[width=.875\linewidth]{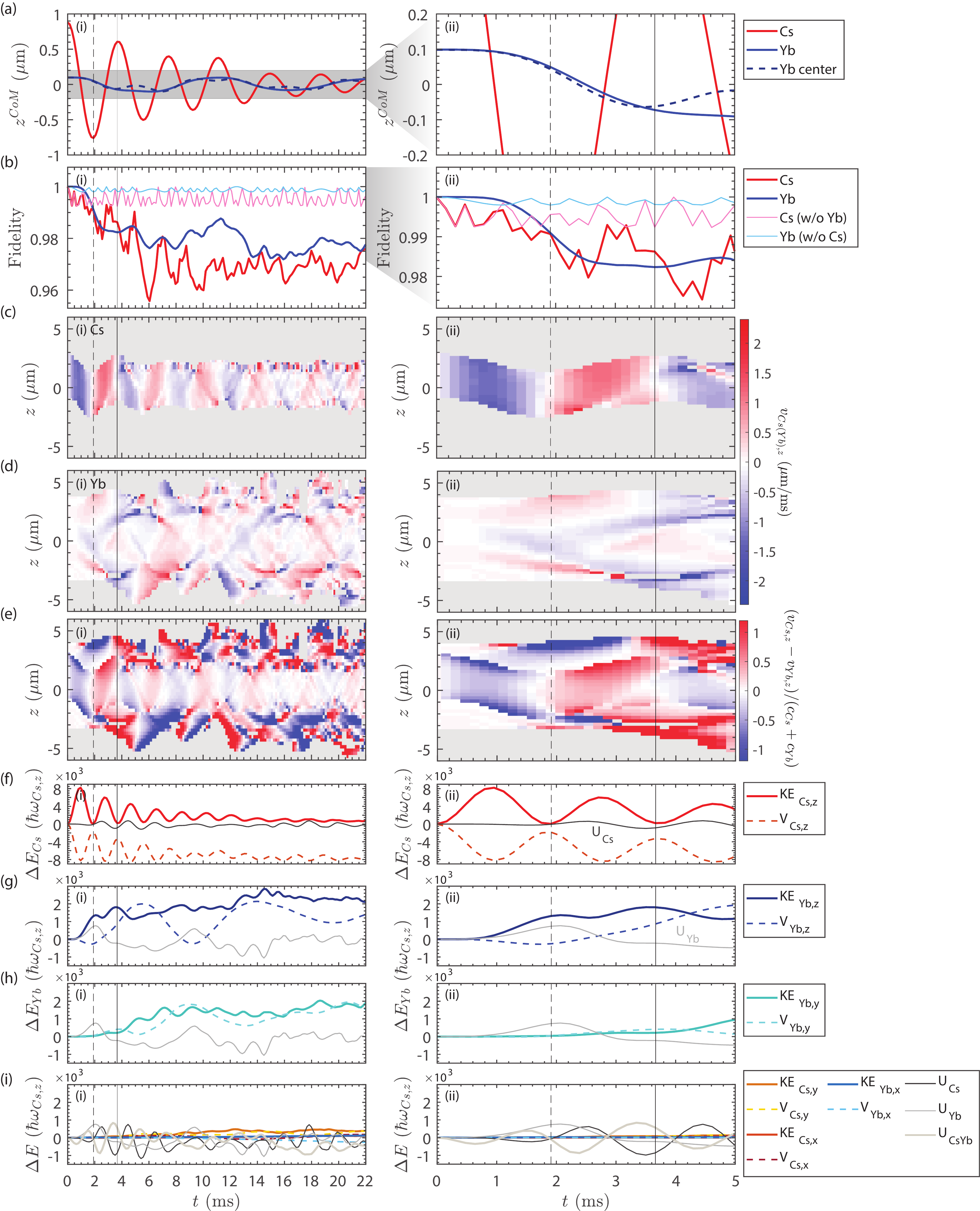}
\caption{Analysis of coupled dynamics during the first 22\,ms (left, (i)) and its corresponding zoomed-in version during the first 5\,ms (right, (ii)). (a) CoM oscillation of Cs (red) and Yb (blue) with the CoM of the central regime of Yb (defined by $|x|<10\;\mu$m, $|y|<6\;\mu$m and $|z|<3\;\mu$m) also shown by the dashed line.
(b) Evolution of fidelity for both Cs and Yb with and without the cross-interaction. 
(c)-(d) Cs and Yb velocity fields along the $z$ axis (defined by $v_\txr{Cs(Yb)}=\hbar\nabla \txr{arg}[\psi_\txr{Cs(Yb)}]/m_\txr{Cs(Yb)}$).
(e) Scaled relative local velocity $(v_\textrm{Cs}-v_\textrm{Yb})/(c_\txr{Cs}+c_\txr{Yb})$ along the $z$ axis, where $c_\txr{Cs(Yb)}=\sqrt{g_\txr{Cs(Yb)}n_\txr{Cs(Yb)}/m_\txr{Cs(Yb)}}$ is the sound speed evaluated by its local density.
(f)-(g) Changes in kinetic and potential energies of Cs and Yb along the $z$ direction.
(h) Changes in kinetic and potential energies of Yb along the $y$ direction.
(i) Corresponding kinetic and potential energy changes of both Cs and Yb in all other directions.
Plots (f)-(i) also display corresponding self-interaction energies for Cs ($U_{\rm Cs}$) and Yb ($U_{\rm Yb}$). The vertical dashed (solid) lines in all plots indicate $T_\txr{Cs}/2~(T_\txr{Cs})$.
}
\label{SM_Fig_2}
\end{figure*}

As discussed in the main text (see also Fig.~2(c) of the main paper) the dominant dynamics, energy transfer and dissipation manifest themselves over the first $\sim 22$\,ms. 
In this subsection we try to shed more light on the complex key dynamical features during this evolution phase.
In order to do this, we present in Fig.~\ref{SM_Fig_2} the study of a  number of relevant observables, both over the first 22\,ms dominant timescale (left panels), and also zoomed in to examine the initial system behaviour over the first 5\,ms (right panels), which enable a more clear visualization over the first complete oscillation cycle, which seeds the subsequent dynamics.

Given that Cs has a smaller spatial extent than Yb along both $z$ and $x$ axes and thus only interacts with the central part of Yb, we can expect minor modifications between the central part of the Yb component (which is co-located with Cs) and the corresponding behaviour of the entire Yb cloud, due to excitations propagating towards the Yb trap edges along the $z$ direction. We can characterize this by looking at the CoM of Yb when measured either as a whole (as done in the main text), or when instead focusing only on the relevant region of direct interactions with Cs.
Specifically, we contrast the earlier findings to the `local' Yb CoM oscillations extracted within the central region
$|x|<10\;\mu$m, $|y|<6\;\mu$m and $|z|<3\;\mu$m.
These features, plotted in Fig.~\ref{SM_Fig_2}~(a), reveal some evidence for secondary (non-central) dynamics in the Yb CoM, as a result of  micro-motions in Yb within and beyond its region of co-existence with Cs. Such micro-motions, which we will show can even feature oscillations in opposing directions for the central and total Yb component, are associated with the emerging excitations.

To determine the difference between the dynamical and initial (static) profiles for each component (which is a measure of the extent of excitations present in such systems~\cite{Pitaevskii2003}), we measure their overlap for each component through the `Fidelity' defined here as
\begin{equation}
\txr{Fidelity}_\txr{Cs(Yb)}\equiv\frac{1}{N_\txr{Cs(Yb)}}\int d\mb{r}\left|\psi_\txr{Cs(Yb)}^{\textrm{ref}}(t)\right|\left|\psi_\txr{Cs(Yb)}(t)\right| \;.
\end{equation}
Here $\psi_\txr{Cs(Yb)}^{\textrm{ref}}\left(\mb{r}(t)\right)=\psi_\txr{Cs(Yb)}\left(\mb{r}-\Delta z^{\textrm{CoM}}_\txr{Cs(Yb)}(t)\hat{\mb{z}},t=0\right)$ is constructed from the initial wavefunction of each component 
[$\psi_\txr{Cs(Yb)}(\mb{r},t=0)$], 
by shifting it along the $z$ direction by an amount dependent on 
its CoM difference 
\begin{equation}
\Delta z^{\textrm{CoM}}_\txr{Cs(Yb)}(t)=z^{\textrm{CoM}}_\txr{Cs(Yb)}(t)-z^{\textrm{CoM}}_\txr{Cs(Yb)}(0), \nonumber
\end{equation}
and computed numerically via interpolation.

Uncoupled oscillations for the two components give a Fidelity $\approx 1$ (with some undamped small amplitude oscillations), as demonstrated by the thin/light pink/blue lines in Fig.~\ref{SM_Fig_2}(b).
Contrary to this, the Fidelity begins to drop rapidly for both components after $\approx 1$\,ms, with an evident few\,\% drop maintained at all subsequent evolution times, clearly demonstrating the impact of incoherent excitation processes during the first coupled oscillation cycle.

We attribute this feature to the time-evolving relative speed between the two components.
At the outset, the two components are moving in the same direction towards decreasing values of $z$, but starting with different offsets and with highly distinct trap frequencies (specifically along the direction of excitation, the frequencies differ by a factor of $\approx 3$), signalling very different oscillation dynamics.
As such, the Cs CoM acquires negative values already around $\approx 1$\,ms (as opposed to $\approx 2.7$\,ms for Yb), with Cs reversing its direction of motion already at $\approx 1.9$\,ms, at which time Yb is still moving along the original (decreasing $z$ direction).
At such times, we are already witnessing  [Fig.~\ref{SM_Fig_2}(a)(ii)] the onset of a small deviation between the central Yb CoM evolution, and that of the whole Yb cloud.
To gain more insight into the coupled oscillations, we also probe [Fig.~\ref{SM_Fig_2}(c)-(d)] the local velocity fields along the $z$ direction for Cs and Yb, respectively. 
These plots include a grey mask to filter out the information for densities lower than $2\;\mu\txr{m}^{-3}$, as this enables us to better highlight the most relevant features.

Initially, and for the first half oscillation cycle (up to $\approx 1.9$\,ms), Cs exhibits coherent motion, with a uniform (but time-dependent) velocity profile.
Contrary to this, the Yb component demonstrates gradual emergence of incoherent features during this early period, in the form of counter-propagating density waves arising from the relative motion between the two components. 
During this timescale, we can thus visualize the initial excitation process as a single-excitation in Yb, caused by the coherent Cs component moving through the central Yb region, which acts much like a red-detuned laser being accelerated through Yb. 

At $\approx1$\,ms, the Cs CoM reaches the trap centre and the coherent motion in Cs reaches the maximum velocity for its harmonic oscillation. At $1.9$\,ms, while Yb is still moving towards lower z values, the Cs CoM reverses its direction, thus creating a direct counterflow between the two components. The magnitude of the counterflow is gradually increased and maintained at high values over the next $\approx$1\,ms when the Cs oscillates to its center for a second time at $2.9$\,ms. This is clearly visible in Fig.~\ref{SM_Fig_2}(e), which depicts the spatial-temporal evolution of the (conveniently scaled) relative velocities.
Here, the local velocities along the $z$ axis for each component $\alpha$ (= Cs, Yb), are obtained from $v_\alpha=\hbar\nabla\txr{arg}[\psi_{\alpha}]/m_\alpha$.
Their relative velocity is shown here scaled to the sum of the instantaneous local speeds of sound, which can be visualized as a potential upper bound for the critical velocity of the coupled system.
This choice is based on the predicted critical velocity of counterflow for two {\em weakly-coupled} superfluids of $(c_\txr{Cs}+c_\txr{Yb})$~\cite{Abad2015,Laurent2019}, where $c_\alpha=\sqrt{g_{\alpha}n_\alpha/m_\alpha}$ with $g_{\alpha}=4\pi\hbar^2a_{\alpha\alpha}/m_\alpha$ is the critical sound speed for each individual component, and the density $n_\alpha$ is measured locally at each moment in time.

The establishment of direct counterflow between the Cs and Yb components leads to the emergence of incoherent features in the propagation of Cs, and coinciding secondary excitations of Yb. 
Such larger relative speeds are generally reached towards the edges of the Cs component, so typically for $|z| \gtrsim 3 \mu$m, suggesting such secondary excitation processes originate at the edges of the Cs component.
These features therefore appear consistent with the generation of coupled excitations in the two components, as a result of the counterflow instability near the edges of the component overlap where the densities, and thus speeds of sound, are appropriately reduced.

Interestingly, at $\approx 3.7$\,ms, when the Cs CoM next reverses its direction, we see a stark difference manifesting itself in the behaviour of the central `local' Yb CoM from the corresponding CoM behaviour of the full Yb system [Fig.~\ref{SM_Fig_2}(a)(ii)]. This indicates the existence of opposing flows in Yb along the $z$ direction, and can be thought of as the extension of co-moving and counter-moving modes to our complicated system of different trap frequencies.
The existence of opposing flows in Yb, in turn gradually couples to the Cs component, leading to a rather complicated excitation pattern after the first few oscillations, as evident from the left panels of Fig.~\ref{SM_Fig_2}(c)-(d).
Such features can also be seen in the subsequent Fig.~\ref{fig_CoM_freq} revealing co-existence of dynamical components for both Cs and Yb across the same dominant frequencies.

Our findings can be reinforced by a detailed analysis of the changes $\Delta E$ in energy exhibited by the various single-particle energy contributions along the different directions [Fig.~\ref{SM_Fig_2}(f)-(i)].
During the period $\approx (2-4)$\,ms, when Cs and Yb are primarily moving in opposing directions, we see the gradual establishment of an additional important mechanism of internal energy transfer within the Yb component, specifically from its directly excited (by the moving Cs component) $z$ direction, to the other transversal $y$ direction of the elongated cloud ($\omega_{{\rm Yb},y} = 1.5 \omega_{{\rm Yb},z}$). This feature is evident in Fig.~\ref{SM_Fig_2}(h)(ii). Such irregularly-oscillating energy exchange across the two tight Yb directions (visible already in Fig.~2(c) of the main paper) saturates after $\approx 20$\,ms,  signalling the end of the primary phase of rapid evolution.

During this period, excitations along the other directions are present [see Fig.~\ref{SM_Fig_2}(i)], but suppressed in comparison, with only modest energy transfer on a much longer timescale (due to the smaller natural frequencies).
This explains why we did not consider energy transfer along such channels as important in the main text.

The above energy calculations across the three directions $x$, $y$, and $z$ (collectively labelled by $\nu$) are based on the single-particle Hamiltonian contributions
\begin{equation}
    E_{\textrm{Cs(Yb)},\nu}^0=\int d\mathbf{r}\psi_{\textrm{Cs(Yb)}}^\ast\hat{H}^0_{\textrm{Cs(Yb)},\nu}\psi_{\textrm{Cs(Yb)}},
\end{equation}
where
\begin{eqnarray}
    \hat{H}^0_{\textrm{Cs(Yb)},\nu} 
    &=& 
    {\rm KE}_{\textrm{Cs(Yb)},\nu} + {\rm V}_{\textrm{Cs(Yb)},\nu}
    \nonumber \\
    &=&
    -\frac{\hbar^2\partial_\nu^2}{2m_{\textrm{Cs(Yb)}}}+\frac{m_{\textrm{Cs(Yb)}}}{2}\omega_{\textrm{Cs(Yb)},\nu}^2 \, \nu^2,
\end{eqnarray}
includes both kinetic energy (KE) and potential energy (V) contributions, with intra- and inter-component interaction energies respectively evaluated by
\begin{equation}
U_{\rm Cs(Yb)}=\frac{g_{\rm Cs(Yb)}}{2}\int d\mathbf{r}\left|\psi_\textrm{\rm Cs(Yb)}\right|^4,
\end{equation}
and
\begin{equation}
U_{\rm CsYb}=g_{\rm CsYb}\int d\mathbf{r}\left|\psi_\textrm{ Cs}\right|^2\left|\psi_\textrm{Yb}\right|^2 \;.
\end{equation}

The abundant excitations in the system begin to gradually thermalize, causing further damping, minimizing the subsequent beat amplitudes, the origin of which is explained in more detail in Sec.~\ref{Sec_beating} below.

\subsubsection{Excitation Spectrum and Origin of Beating \label{Sec_beating} }

\begin{figure}[b!]
\centering
\includegraphics[width=1\linewidth]{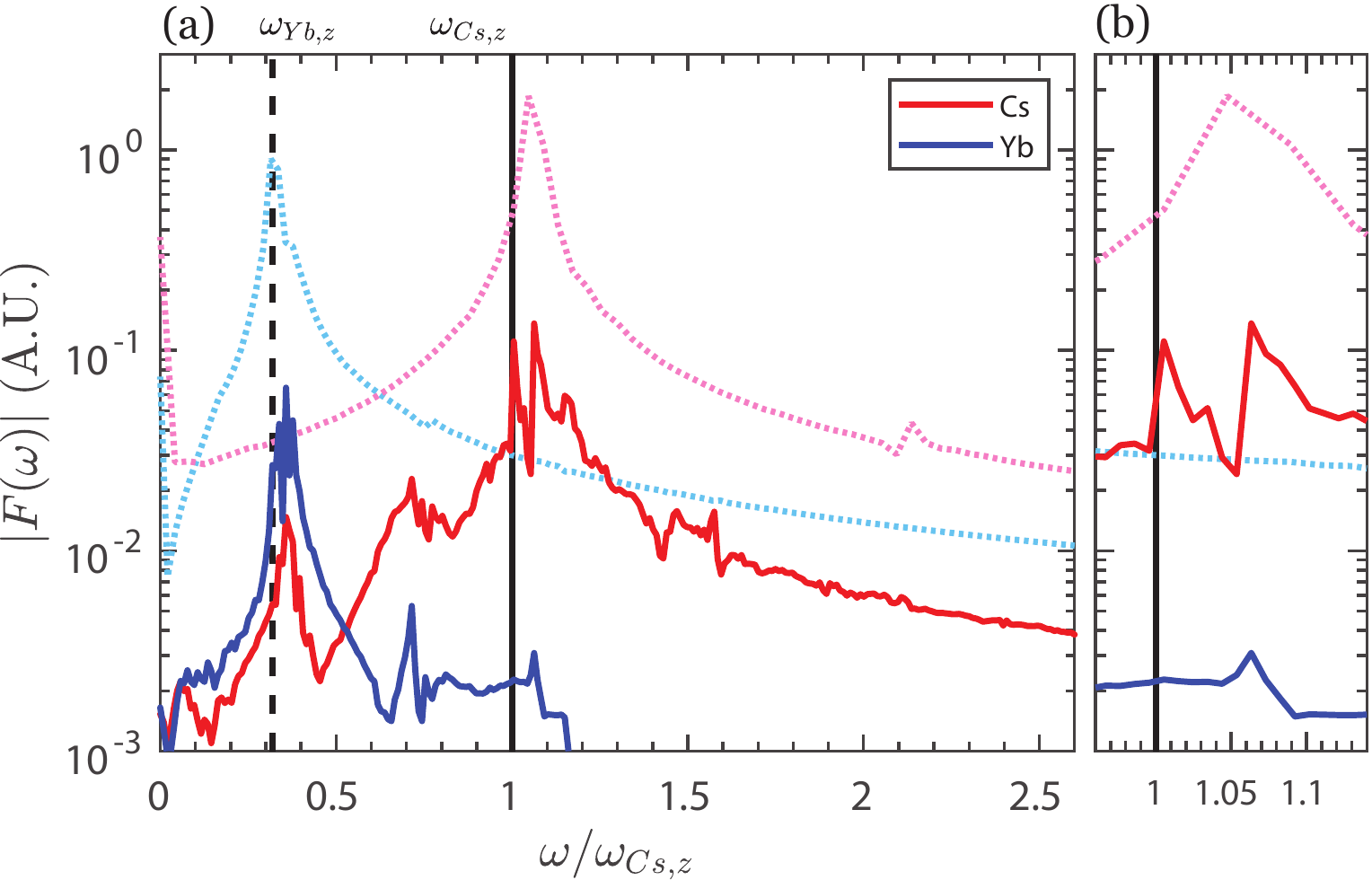}
\caption{The Fourier spectra of CoMs for $(N_\textrm{Yb},N_\textrm{Cs})=(40,6.5)\times10^3$ (based on an evolution of $650/\omega_{\txr{Cs},z}\approx413.8$~ms) The red and blue solid lines are the spectra of Cs and Yb, respectively. The pink and blue dotted lines are the corresponding spectra for the case where the other component is static. In (b) we zoom-in to the region around $\omega / \omega_{\textrm{Cs},z}=1$ to show the splitting of the oscillation frequency in Cs.}
\label{fig_CoM_freq}
\end{figure}

The frequency shift of dynamical collective modes of coupled superfluids due to the cross interaction between components has been studied in the context of both Bose-Bose and Bose-Fermi superfluid mixtures~\cite{Miyakawa2000,Ferrier2014,Laurent2019,Asano2020a,Asano2020b}.
In the Cs-Yb mixture under consideration here, the Cs particle number is much smaller than the Yb number. We therefore may neglect the back action of the Cs on the Yb along the $z$ axis. As a consequence, the Yb component oscillates with the harmonic trap frequency $\omega_{\txr{Yb},z}$. Moreover, within the Thomas-Fermi approximation, the Yb density profile, given by $n_\txr{Yb}(z)=\left[\mu_\txr{Yb}-V_{\txr{Yb},z}(z)\right]/g_{\txr{Yb}}$, can be used to estimate the effective potential acting on Cs due to interspecies interactions.
The effective Cs potential along the $z$ axis is then $V_{\txr{Cs},z}^{\txr{eff}}(z)\approx V_{\txr{Cs},z}+({4\pi\hbar^2a_\textrm{CsYb}}/{M_\mu})n_\txr{Yb}(z)$. This potential has an effective harmonic frequency of 
\begin{equation}
\omega_{\txr{Cs},z}^{\txr{eff}}=\omega_{\txr{Cs},z}\sqrt{1-\frac{a_\txr{CsYb}m^2_{\txr{Yb}}}{a_\txr{Yb}M_\mu m_\txr{Cs}}\frac{\omega_{\txr{Yb},z}^2}{\omega_{\txr{Cs},z}^2}}.
\end{equation}
This simple formula predicts $\omega_{\txr{Cs},z}^{\txr{eff}}\approx1.05\omega_{\txr{Cs},z}$, which captures the effective frequency for the Cs in the presence of a static Yb cloud in the following discussion.

The previous subsection studied the dominant dynamical regime of the first $\approx22$\,ms, where most of the Cs excess energy is transferred to Yb through the generation of excitations due to their relative motion.
Detailed numerical consideration of the heavily damped coupled oscillations indicates both the existence of multiple dynamical modes, and the emergence of beating on longer timescales.
Here we shed more light on both these features, and the dominant oscillation frequencies of the coupled mixture, by analyzing the mixture's excitation spectrum, through the Fourier spectra of $z_\txr{Cs(Yb)}^{\txr{CoM}}$.
These are shown by the red (Cs) and blue (Yb) solid lines in Fig.~\ref{fig_CoM_freq}.

We observe a multitude of excitation frequencies in both components, consistent with the complicated excitation generation discussed above.
As expected, the dominant dynamics for each component arise at  frequencies close to their natural frequency, slightly upshifted due to the mutual attractive mean-field potential, as discussed above.
Such peaks, and the general underlying structure for each component, reflect well the excitation spectrum that would be expected when considering the modification imposed by the attractive mean-field potential, but ignoring the dynamical back action of the other component.
To better characterize such an effect of the attractive mean-field potential, we have also performed corresponding dynamical numerical simulations for Cs (and Yb) when coupled to the static other component, i.e. respectively Yb (and Cs). This has been done by adding to our simulations a {\em static} mean-field contribution from the second component, set to its configuration at $t=0$ in our simulations, whose role is to change the effective (trap + mean field) potential felt by the main component. Such results are shown for the Cs (and Yb) component by the pink (blue) dotted line, clearly revealing the previously mentioned $\approx 5\%$ frequency upshift for Cs (and the comparatively smaller upshift for the more massive Yb component, which is affected less by the Cs attraction).

Beyond the dominant contributions for each component discussed above, the dynamical excitation frequencies reveal a multitude of other relevant modes.
Firstly, we find each component to be additionally responding (in their central overlap region) to 
the mutual attractive mean field by co-moving at the dominant frequency of the other superfluid. In other words, parts of the (centrally-located) Yb are dragged along by Cs at $\approx \omega_{\txr{Cs},z}$, while correspondingly there is a notable component of Cs being dragged along by Yb at $\approx \omega_{\txr{Yb},z}$. 
We also find a clear (broad) peak for both components around their frequency difference $\approx (\omega_{\txr{Cs},z} - \omega_{\txr{Yb},z}) \approx 0.7 \omega_{\txr{Cs},z}$ [which is also close to the average frequency $(\omega_{\txr{Cs},z} + \omega_{\txr{Yb},z})/2$].
The simultaneous presence of dominant oscillation modes for the two components both close to the natural frequencies of each component, and close to the difference (and average value) of such frequencies, indicates the strong presence of co-moving and counter-moving oscillating components. Here the dynamics are complicated by the different masses, atom numbers, interaction strengths and trapping potentials for the two components.

Beyond such modes, our spectra clearly demonstrate the emergence of splitting of the dominant upshifted frequencies of each component into two (or multiple) close-by frequency peaks.
Such frequencies are generally found to be upshifted from the natural frequency, and to lie on either side of the static-mean-field contribution, thus highlighting the dynamical role of the other component.
Specifically for Cs, we find two dominant frequencies, of broadly comparable importance.
For the `primary' atom number combination considered here, these are 
$\omega_{\txr{Cs},z}^{(1)} \sim 1.063\omega_{\txr{Cs},z}$ and 
$\omega_{\txr{Cs},z}^{(2)} \sim 1.005\omega_{\txr{Cs},z}$.
The difference in these two frequencies establishes a timescale 
$2 \pi /\left|\omega_{\txr{Cs},z}^{(1)} - \omega_{\txr{Cs},z}^{(2)}\right|^{-1} \approx 69$\,ms which leads to the underlying beating cycle reported in Fig.~2(c) of the main text. The period of this beating cycle is highly sensitive to the particular atom numbers considered. Additional features seen within each beating cycle can be attributed to the intricate interplay between the various relevant modes discussed above, which will in general have different relative phases and damping rates, and lead to dynamics on a faster, few ms, timescale.

\begin{figure*}[t!]
\centering
\includegraphics[width=1\linewidth]{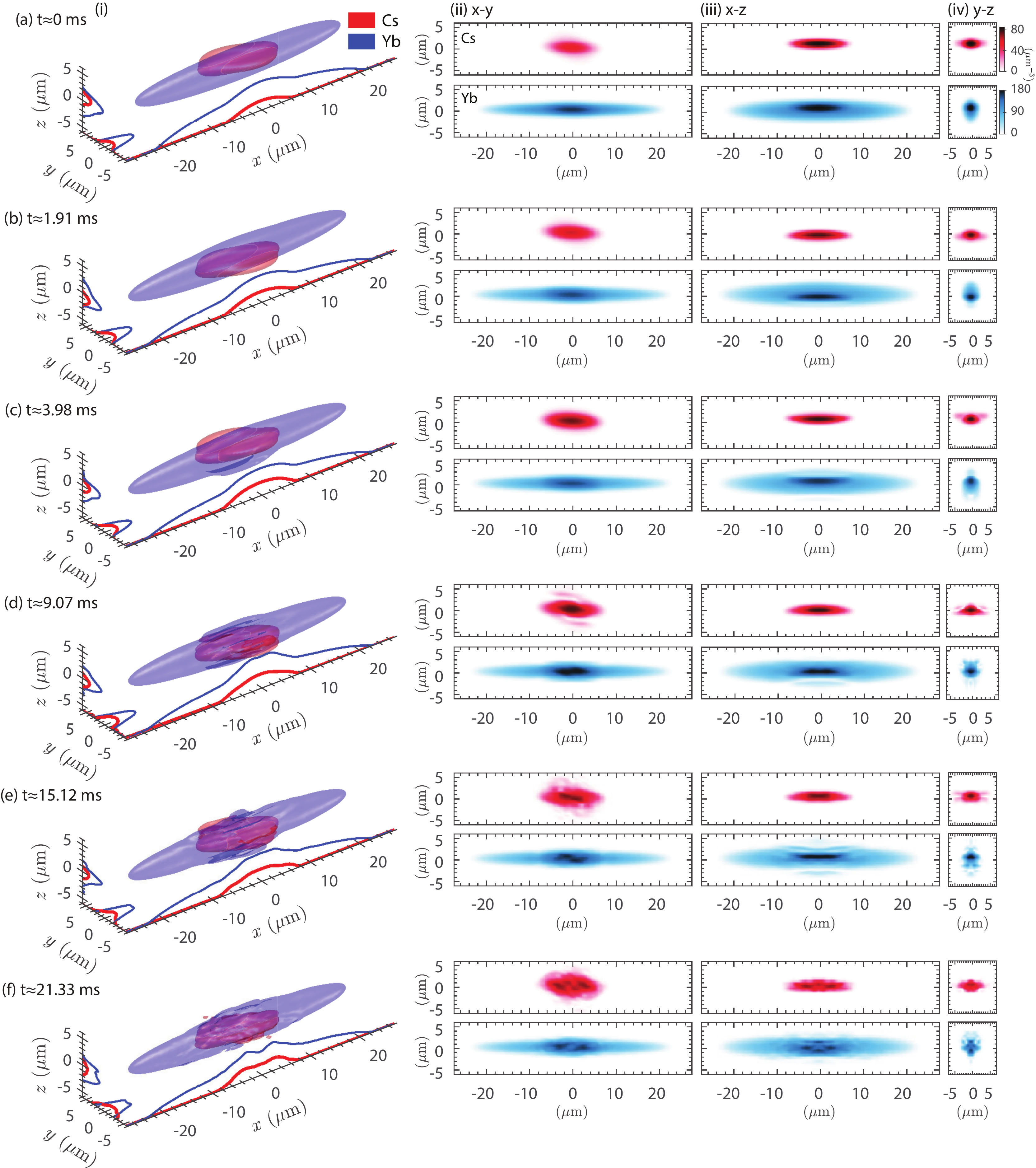}
\caption{
Density snapshots of Cs (red) and Yb (blue) at different times during the first $\approx21$\,ms of coupled evolution.
Shown are (i) 3D profiles corresponding to density isosurfaces for the 5$\%$ peak density of Yb at $t=0$ with corresponding slices shown along each axis and 2D sliced density snapshots across the (ii) $x-y$ ($z=0$) plane, (iii) $x-z$ ($y=0$) plane and (iv) $y-z$ ($x=0$) plane.
\label{SM_Fig_4}
}
\end{figure*}

\subsection{Density Snapshots}

To better visualize the complicated dynamics, we present in Fig.~\ref{SM_Fig_4} representative 3D density isosurfaces and 2D sliced density snapshots during the first $\approx22$\,ms of coupled evolution, showing also the corresponding 1D slices. The full evolution, which also includes later times can be found in the accompanying Supplementary Video.
From these, we can see the different stages of the evolution, including the Yb component being dragged by Cs over the first $\approx 2$\,ms, the subsequent double-peak distribution of Yb along the $z$ axis, and the gradual excitations along different directions.

\section{Collapse Properties and Dynamics}
We model the experiment using a three-dimensional mean-field description,
expressed mathematically as two-component coupled GPEs with an added three-body loss term \cite{Altin2011} for Cs as detailed in Eqs.~\ref{eq:GPE_Yb}-\ref{eq:V_Cs}.  Here we have $\Delta z = 0$, and $\omega_\mathrm{z} = 2\pi \times 260\,\mathrm{Hz}$. We take $a_{\mathrm{Yb}} = 105\,a_0$ and $a_{\mathrm{CsYb}} = -75 \,a_0$.
Our numerical simulations of these coupled GPEs, described below, use Fourier pseudospectral methods with adaptive Runge-Kutta timestepping and are implemented using the XMDS2 software \cite{Dennis2013}.

\subsection{Ground states and static collapse analysis}
One can attempt to find ground state solutions [$\partial_t \psi_{\mathrm{Yb}} = \partial_t \psi_{\mathrm{Cs}} = 0$] to the coupled GPEs with a fixed Cs scattering
length and zero three-body loss $K_{3,\mathrm{Cs}}^{(\mathrm{C})}=0$. Below 
a critical value of Cs
scattering length, $a_{\mathrm{Cs}}^{(\mathrm{crit})}$, collapse
occurs~\cite{Donley2001a} and mean-field ground state solutions cease to exist.
We use imaginary time propagation to find ground state solutions numerically.
For Cs only, with $N_{\mathrm{Cs}} = 4 \times 10^3$, we find
$a_{\mathrm{Cs}}^{(\mathrm{crit})} \approx -2.15\,a_0$. We found that an
approximate variational calculation \cite{parker_etal_jpb_2007,
  billam_etal_variational_2011} assuming a Gaussian ground state profile (and ignoring rotation of trap axes) yields
a fairly similar result ($a_{\mathrm{Cs}}^{(\mathrm{crit})} \approx -2.48\,a_0$) without the computational effort of solving the GPE. With
Yb also present, with atom numbers $N_{\mathrm{Cs}} = 4 \times 10^3$ and
$N_{\mathrm{Yb}} = 5 \times 10^4$, we find $a_{\mathrm{Cs}}^{(\mathrm{crit})}
\approx 51.25\,a_0$ using imaginary time propagation. In this case, a Gaussian variational calculation gives
incorrect results ($a_{\mathrm{Cs}}^{(\mathrm{crit})}
\approx -1.75\,a_0$) since the Yb condensate is far from having a Gaussian profile.

\subsection{Dynamical collapse analysis}
To model the dynamic collapse experiments reported in the main text, we perform
real-time simulations of the coupled GPEs that are intended to match the
experimental protocol. We begin from the Cs+Yb ground state solution, found as
described above, with $a_{\mathrm{Cs}}(t=0) = 147\,a_0$, $N_{\mathrm{Cs}} = 4
\times 10^3$, and $N_{\mathrm{Yb}} = 5 \times 10^4$. When simulating the Cs-only
dynamic collapse experiments we continue to find the Cs+Yb ground state, but we
remove Yb atoms immediately at time $t=0$ by setting $\psi_{\mathrm{Yb}} =0$, in
analogy to the experimental protocol.

We then perform a real time simulation of the coupled GPEs in which the Cs
three-body loss coefficient is set to a constant, non-zero, value and the Cs
scattering length is varied in time to match the experimental protocol.
Specifically, we implement in sequence a hold at $a_{\mathrm{Cs}}(t) = 147\,a_0$ for $\SI{5}{m s}$, a $\SI{10}{m s}$ linear ramp to $a_{\mathrm{Cs}}(t) =
a_{\mathrm{Cs}}^{(\mathrm{jump})}$, a hold at this value for $\SI{30}{m s}$, a
$\SI{10}{m s}$ ramp back to $a_{\mathrm{Cs}}(t) = 147\,a_0$, and a final
$\SI{1}{m s}$ hold at this value. At this time in the experimental protocol a
time-of-flight measurement begins; in the simulations we simply read out the
final atom numbers. We repeat this process for varying values of
$a_{\mathrm{Cs}}^{(\mathrm{jump})}$, obtaining the results reported in the
main text. All parameters are set to experimentally measured values rather than
fitted to the data, except for the Cs three-body loss coefficient. The latter is
less precisely constrained by experimental measurements. In practice we repeated
the simulations for a few trial values and found that
$K_{3,\mathrm{Cs}} = \SI{1.8e-27}{cm^6 s^{-1}}$ gives a
reasonable quantitative match to the results of the collapse experiment, while
being consistent with other experimental measurements \cite{Kraemer2006,Guttridge2017}. We have not searched for a
precise best-fit value of $K_{3,\mathrm{Cs}}^{(\mathrm{T})}$ (e.g., by
least-squares fitting) because of the long computational times needed for the simulations.

By default our data points are from simulations in a box of dimension $(L_x, L_y,L_z)$ = $(96, 24, 24)\,
\SI{}{\micro m}$ with grid resolution of $3/2$ points per $\SI{}{\micro m}$ in
each direction. We checked convergence of the final $N_{\mathrm{Cs}}$ value to
$\approx 1\%$ by repeating selected simulations at higher resolution ($3$ points
per $\SI{}{\micro m}$) and in larger boxes in the most-constrained direction
($L_y = \SI{48}{\micro m}$). Our data points with the lowest $a_{\mathrm{Cs}}^{(\mathrm{jump})}$ value in each simulation were obtained from higher-resolution simulations, for which the final $N_{\mathrm{Cs}}$ value agreed to within $\approx 1 \%$ with the larger-box simulations. For values of $a_{\mathrm{Cs}}^{(\mathrm{jump})}
\lesssim -0.7 \,a_0$ (Cs-only), or $\lesssim 51\, a_0$ (Cs+Yb), the three-body
loss coefficient is not sufficient to prevent a mean-field collapse, even in our
higher-resolution simulations. When this occurs the final Cs number is not
converged and changes with the grid resolution. Quantitative modeling of this
collapse regime would require a beyond-mean-field description, which is beyond
the scope of this work.

\end{document}